# A Stackelberg Game Theoretic Model of Lane-Merging

Jehong Yoo, *Member, IEEE* and Reza Langari, Senior *Member, IEEE*

**Abstract— Merging in the form of a mandatory lane-change is an important issue in transportation research. Even when safely completed, merging may disturb the mainline traffic and reduce the efficiency or capacity of the roadway. In this paper, we consider a Stackelberg game-theoretic driver behavior model where the so-called utilities or *payoffs* reflect the merging vehicle's *aggressiveness* as it pertains the decision to merge as the situation stands or to accelerate/decelerate prior to the actual lane-change maneuver. The interaction of the merging vehicle with the mainline traffic is also considered whereby the combination of aggressiveness of the respective vehicles leads to both longitudinal and lateral disturbances to the mainline flow as well as subsequent reduction in the roadway throughput. The present study shows in semi-quantitative form that this impact depends on the level of aggressiveness of the merging and mainline vehicles in an intuitive manner, leading to the potential use of this model in traffic flow analysis and autonomous driving.**

## I. INTRODUCTION

Roadway congestion impacts the quality of life in metropolitan areas and as such is a major concern in traffic engineering. A variety of factors contribute to roadway congestion but improper lane-changing, be it in discretionary mode or in the mandatory case, as in merging onto the mainline of a highway, is one of the key factors that impacts roadway congestion by disturbing the mainline traffic flow. In the worst case, improper merging causes accidents that, aside from their unfortunate impact on those immediately involved, can have drastic impact on traffic. Improper merging itself has multiple causes, including driver inexperience or inattentiveness to the roadway configuration and traffic setting.

In particular, willful or inherent *aggressiveness* is known to lead to improper merging behavior [1]. However, few studies have quantified the impact of aggressiveness (or related factors) on merging and its overall impact on traffic flow. To be sure, understanding human driving has been a major area of study in traffic engineering [2-8]. Several models have indeed been developed to delineate the human driving behavior [9-14], and which can be classified into *longitudinal* and *lateral* models.

The latter, in particular, reveal the heuristics that characterize a driver's lane change or merging behavior [15-18]. An aggressive disposition is a notable factor in this respect with the extreme case of road rage often cited as a particular threat to traffic safety [19, 20]. To this end, and as part of a longer study, we have proposed a highway driving model [21] that uses Stackelberg game theory [22] to model drive behavior in discretionary lane changes as a function of driver aggressiveness. Ahmed [23], among others, has utilized a gap acceptance model for discretionary lane-changing purposes. Likewise, Hidas [24] has proposed the notion of *driver courtesy*, a type of cooperation among drivers, in modeling lane-change and merging operations. Kim [25] has proposed a modified IDM (Intelligent Drier Model) to deal with conflicts among a group of vehicles and is, in this respect, similar to Hidas's driver courtesy scheme.

In addition, Swaroop and Yoon have developed an emergency lane-change maneuver within the overall framework of vehicle platooning [26] while Jula et al. have considered a minimum longitudinal spacing criterion to avoid crashes [27]. Kanaris and Ioannou have also proposed a certain minimum safety spacing for lane-changing and merging in automated highway systems [28]. Hall and Tsao [29] studied merging algorithms for autonomous vehicles, cooperative vehicles, and cooperative platooning [30]. Mine and Mimura studied the merging model with an acceleration area in relation to the probability density function of the delay in the mainline traffic flow [31]. Wang and et al. studied a proactive merging algorithm that adjusts the velocity of the ramp car prior to merging [32].

On the other hand, Bose and Ioannou [33] showed that the disturbance from merging to the mainline traffic can be analyzed using a linear car following model (as for instance discussed by Pipes [34].) In particular, it is well known that a small delay at one point can lead to severe congestion upstream, often referred to as the *slinky* effect or string instability [35]. We will show later in this work that this initial disturbance is potentially the result of certain combinations of driving

---

This work was partially supported by Qatar National Research Fund, NPRP-09-393-2-145.

J.-H. Yoo was with Texas A&M University, College Station, TX 77843 (email: jehong.yoo@tamu.edu.)

R. Langari is with Texas A&M University, College Station, TX 77843 (e-mail: rlangari@tamu.edu).

behaviors that may originate in driver aggressiveness (including an overly cautious disposition) during the merging operation.

### B. Game Theory in Transportation

Game theory has been applied in various ways to study the effects of policy, decisions, and/or the actions of individual agents in a transportation system. These studies can be broadly classified into two categories: infrastructural regulation studies (traffic control problem) and agent-oriented studies (vehicle placement or route decisions). In the first category, researchers have employed game theory in relation to dynamic traffic control or assignment problems. For instance, Chen and Ben-Akiva [36] adopted a non-cooperative game model to study the interaction between a traffic regulation system and traffic flow to optimally regulate the flow on a highway or an intersection while Zheng-Long and De-Wang [37] addressed the ramp-metering problem via Stackelberg game theory. Su et al. [38] have also used game theory to simulate the evolution of a traffic network.

In the second category, vehicles are regarded as game participants and traffic rules are generally considered to be implicit in the respective decision models [39-41]. In this context Kita [42] has worked to address the merging-giveway interaction between a through car and a merging car as a two-person non-zero sum game. This approach can be regarded as a game theoretic interpretation of Hidas' *driver courtesy* scheme [43] from the viewpoint that the vehicles share the payoffs or heuristics of the lane-changing process. This leads to a reasonable traffic model although the approach does not address the uncertainties resulting from the actions of other drivers. In recent studies, Talebpour et al [44, 45] have considered the notion of *incomplete information* as part of the game formulation process and have developed a model that in certain respects addresses the aforementioned concern. Likewise, Altendorf and Flemisch flow [46] have developed a game theoretic model that addresses the issue of *risk-taking* by drivers and its impact on the traffic flow. Their study focuses on the cognitive aspects of this decision making process and its impact on traffic safety. Likewise, Wang, et al. [47] have used a *differential game* based controller to control a given vehicle's car-following and lane-changing behavior while in [48], the authors have applied an Iterative Snow-Drift (ISD) game on *cross-a-crossing* scenario. In [49], Stackelberg game theory was used solve conflicts in shared space zones while in [50] the authors compared heavy vehicle and passenger car lane-changing maneuvers on arterial roads and freeways. Elvik [51] offers a rather complete review of related works in this area, albeit up to 2014.

### C. Outline of the Current Work

In view of the above remarks, and in the same context as the cited works, we develop a merging model via Stackelberg game theory. In particular, we focus on developing a model with the utilities or *payoffs* (in game-theoretic language) that originate from the drivers' intentions and disposition. This model will determine the instant to merge and the related acceleration/deceleration behavior in view of the driver's *aggressiveness*. The model is intended to simulate a spectrum of driver aggressiveness levels during highway merging maneuvers. We base this merging model on a highway driving model, which previously included discretionary lane-changing [21]. The extension to highway merging (alternatively *mandatory* lane-changing) is nontrivial, however, since it requires active consideration of multiple simultaneous games that involve the merging vehicle and alternate competing vehicles in the mainline traffic. This is in turn due to the fact that in a typical merging situation, the merging vehicle cannot be certain as to whether to accelerate and precede the immediately competing vehicle in the mainline or else await the passing of that vehicle and position itself behind that vehicle (and ahead of the next competing vehicle). In the sequel, this point is discussed at more length and its impact in terms of the longitudinal and lateral disturbances to the mainline traffic is elaborated.

The paper is organized as follows: Section II describes the system configurations for the problem. Section III defines the game theoretic model for highway merging while the related case studies are presented in Section IV. The impact of the driver disposition (aggressiveness) in merging and its consequences on the mainline traffic is presented in Section V. Section VI presents a summary of the work and concludes the paper.

## II. System Configuration

In order to design and simulate a merging model, we endow a given vehicle with three functions: *vicinity recognition*, *decision making*, and *manipulation of the vehicle* as depicted in Figure 1. The vicinity recognition focuses on the information concerning the vehicles within the immediate environment of the subject vehicle. Based on this information, the subject vehicle (or the driver in it) makes a decision, and the decision is implemented in view of the vehicle dynamic model. This same principle is applicable to every vehicle involved in the simulation studies discussed below.

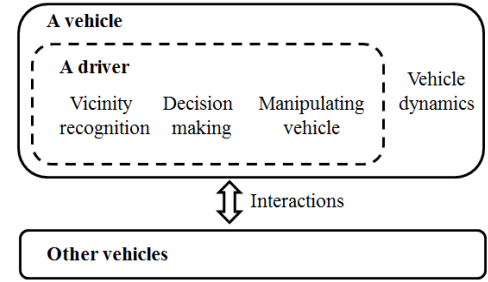

**Figure 1. System Configuration.**

### A. Vehicle Dynamic Model

A two-wheel vehicle model [52] given in (1.1) and the kinematic relations in (1.2) is schematically depicted in Figure 2. This model describes the vehicle pose $(x, y, \theta)$ according to its velocity, $v$, and steering angle, $\delta$, which are determined by the vehicle manipulation controllers:

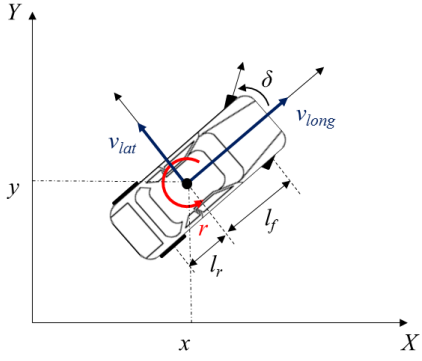

**Figure 2. Planar view of vehicle.**

$$\begin{bmatrix} \dot{v}_{lat} \\ \dot{r} \end{bmatrix} = \begin{bmatrix} -\dfrac{C_{af}+C_{ar}}{mv_{long}} & -\dfrac{l_f C_{af}+l_r C_{ar}}{mv_{long}}-v_{long} \\ -\dfrac{l_f C_{af}-l_r C_{ar}}{I_z v_{long}} & -\dfrac{l_f^2 C_{af}+l_r^2 C_{ar}}{I_z v_{long}} \end{bmatrix}\begin{bmatrix} v_{lat} \\ r \end{bmatrix} + \begin{bmatrix} \dfrac{C_{af}}{m} \\ \dfrac{l_f C_{af}}{I_z} \end{bmatrix}\delta \quad (1.1)$$

$$\begin{bmatrix} \dot{x} \\ \dot{y} \\ \dot{\theta} \end{bmatrix} = \begin{bmatrix} v \cdot \cos\theta \\ v \cdot \sin\theta \\ r \end{bmatrix} \quad (1.2)$$

where $\theta$ denotes the heading angle of the vehicle, $(x,y)$ the position of the vehicle in the Cartesian coordinate system shown in the figure, $r$ the yaw velocity, $v_{lat}$ the lateral velocity, $m$ the mass of the vehicle, $I_z$ its moment of inertia, $v_{long}$ its longitudinal velocity, $v_{lat}$ its lateral velocity, $l_f$ the distance between the front wheel and the center of mass of the vehicle, $l_r$ the distance between the rear wheel and its center of mass, $C_{af}$ its front cornering stiffness, and $C_{ar}$ its rear cornering stiffness.

### B. Driver's Manipulation of the Vehicle

To generate proper acceleration/deceleration and steering angle, we use two independent controllers for longitudinal and lateral control of the vehicle in a manner that we believe to be consistent with human driving. Both controllers (Proportional plus Derivative or PD) are used to control the relative velocity or the headway and the steering angle. The longitudinal control output, the resultant acceleration, is determined by the weighted mean of the two controls of the relative velocity and the headway. Both control outputs are limited by physical constraints of the vehicle and the driver's *disposition*. In particular, an aggressive driver will move more drastically in the longitudinal and lateral directions than a normal or cautious driver. Thus, the vehicle acceleration $a$ is given by

$$a = \min(K_{pg} \cdot e_{v,d} + K_{dg} \cdot \dot{e}_{v,d}, g_l, g_{pl}) \quad (1.3)$$

where $e_{v,d}$ is the error between the reference velocity and the velocity of the vehicle or the error between the reference relative distance and the relative distance between the given vehicle and the vehicle ahead, $K_{pg}$ is the proportional gain of the longitudinal controller, $K_{dg}$ is the derivative gain of the longitudinal controller, $g_l$ is the acceleration limit that can be changed by the driver's disposition, and $g_{pl}$ is the physical limitation of acceleration or deceleration. Likewise, the steering angle is defined by

$$\delta = \min(K_{pl} \cdot e_{lat} + K_{dl} \cdot \dot{e}_{lat}, \delta_{lat}, \delta_{pl}) \quad (1.4)$$

where $e_{lat}$ is the error between the reference lateral position and the lateral position of the vehicle, $K_{pl}$ is the proportional gain of the lateral controller, $K_{dl}$ is the derivative gain of the longitudinal controller, $\delta_{lat}$ is the steering angle limit that can be altered by the driver's disposition, and $\delta_{pl}$ is the physical limitation of the steering angle. Here, $\delta_{lat}$ can be obtained by using the following ratio which is known as lateral acceleration gain [53]:

$$\frac{a_{yl}}{\delta_{lat}} = \frac{v^2}{57.3Lg + K_{us}v^2} \quad (1.5)$$

where $a_{yl}$ denotes the lateral acceleration limit that can be changed by the driver's disposition, $L$ is the wheelbase, $K_{us}$ is the understeer gradient of the vehicle, and $g$ is the acceleration of gravity.

### C. Vicinity Recognition and Characterization

#### 1. Characterization of surrounding vehicles

This function first classifies the given vehicle's surroundings into vehicles *ahead* and those *behind* according to their longitudinal positions with respect to each lane. Here, the vehicles ahead and behind are called the *leading* and *following* vehicles, respectively. Then the nearest vehicles in each lane are chosen as the *vicinity* vehicles as depicted in a generic form[1] in Figure 3, where ① designates the vehicle itself, ② the leading vehicle in Lane 1, ③ the following vehicle in Lane 1, ④ the leading vehicle in Lane 2, ⑤ the leading vehicle in Lane 3, and ⑥ the following vehicle in Lane 3. We will add artificial errors to approximate the uncertainty in recognizing the surroundings according to the driver's disposition as described below.

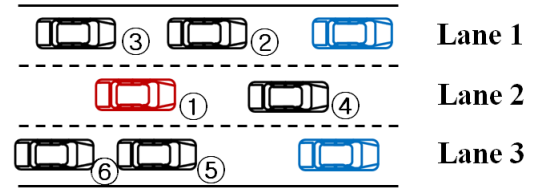

**Figure 3. Vicinity vehicles in generic form.**

#### 2. Driver reaction consideration

Driver reaction time and poor prediction of other vehicles' actions are crucial factors in transportation [54-57]. To incorporate these in our study, we consider the delayed recognition of the presence of other vehicles and late reaction to their presence. In particular, we consider the decision maker's recognition point that is used to identify the lanes occupied by various vehicles as depicted in Panel (a) of Figure 4. It is assumed that the magnification ratio of the rectangle surrounding the intruding vehicle varies from 1 to $1.3^2$ as a function of driver aggressiveness, $q \in [0,1]$, to be elaborated shortly, as depicted in Panel (b) of the figure.

---

[1] This diagram shows the generic configuration for vicinity recognition and will be specialized to the merging situation shortly.

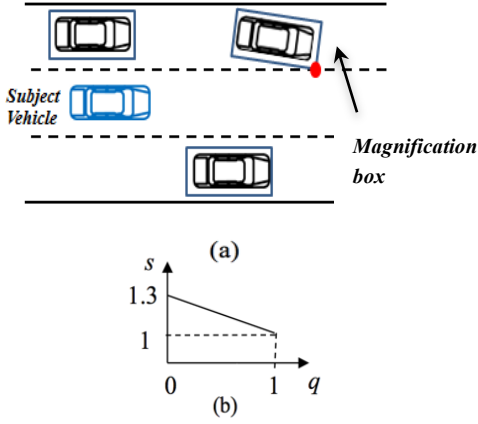

**Figure 4. Boundary recognition response.**

### 3. Collision detection

Vehicles are assumed to be rectangles that have certain widths and lengths. Thus, collision between any two vehicles can be detected in terms of the overlapping area between the projections of these rectangles using the Separating Axis Theorem [58] as depicted in Figure 5.

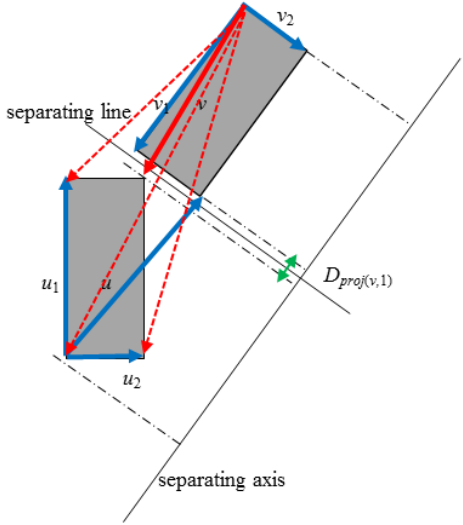

**Figure 5. Schematic view of Separating Axis Theorem.**

To formulate an index $I_{col} \in [0,1]$ to represent the collision possibility, we use the gaps between two rectangles along the separating axis.

$$D_{proj(v,i)} = \begin{cases} \min\left(\left|\dfrac{v_i \cdot v}{v_i}\right|\right) & if \quad \forall \dfrac{v_i \cdot v}{|v_i|} < 0 \\ \min\left(\left|\dfrac{v_i \cdot v}{v_i}\right| - |v_i|\right) & if \quad \forall \dfrac{v_i \cdot v}{|v_i|} > |v_i| \\ 0 & otherwise \end{cases} \quad (1.6)$$

where $D_{proj}$ denotes the gaps along the separating axis, $v_i$ a vector defining the rectangle, and $v$ the vector to the opposite corner. We define the collision possibility index, $I_{col} \in [0,1]$, as the exponential inverse of the gap.

$$I_{col} = e^{-\sqrt{\frac{D_{col,v}^2 + D_{col,u}^2}{2}}} \quad (1.7)$$

where

$$D_{col,v} = \sqrt{D_{proj(v,1)}^2 + D_{proj(v,2)}^2} \,. \quad (1.8)$$

$D_{col,u}$ is defined in a similar manner. Note that the collision possibility index is 1 when two rectangles overlap and 0 when the gap between them approaches infinity.

## III. GAME FORMULATION

With the above in mind, we configure a straight road with one merging lane as depicted in Figure 6. The assumption is that the entry ramp (or the merge lane) has an entrance/acceleration area for merging, and can be viewed as an additional lane on the roadway that only the merging vehicles occupy[3]. In the mainline, there are two kinds of vehicles: vehicles with decision makers and vehicles that just move at preset speeds (not all the same) without changing lanes. The latter (shown in blue) form the boundaries of the simulation.

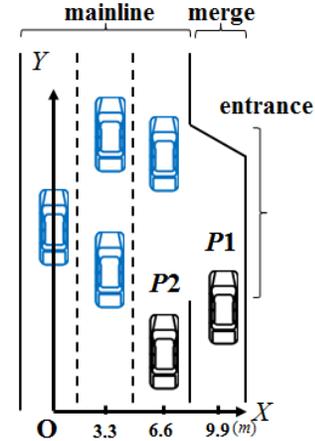

**Figure 6. Problem space.**

Figure 1. Game Definition

We assume a typical merging situation and first establish a

---

Stackelberg game between the merging vehicle (*P*1) and the immediately preceding (or *competing*) vehicle in the mainline (*P*2) as depicted in Figure 6. This happens in a situation when the merging vehicle, *P*1, is generally well positioned to maneuver into the mainline although even in this circumstance its *relative longitudinal position* with respect to the mainline vehicle, *P*2, and their *relative speeds* impacts the decision by *P*1 whether to proceed with the merging manuever. This decision requires a game-therotic analysis that considers the highlighted factors along with aggressivenss of each vehicle in terms of a set of *payoffs* or *utilities* that are defined shortly.

In general, the decision to merge is not as straighforward as it may appear from the above discussion and the diagram in Figure 6. It is entirely possible, and even likley that the merging vehicle would consider options other than to merge immediately. For instance, depending on its own aggressivenss or the perceived threat from the competing vehicle in the mainline, *P*2, it may decide to accelerate to move well ahead of *P*2, possibly positioning itself even ahead of the vehicle that leads *P*2 in the mainline, i.e. *P*2′ as depicted in Figure 7 or declerating to merge behind *P*2 in which case it will compete with vehicle immediately behind *P*2′, which is also denoted by *P*2′ in the figure[4].

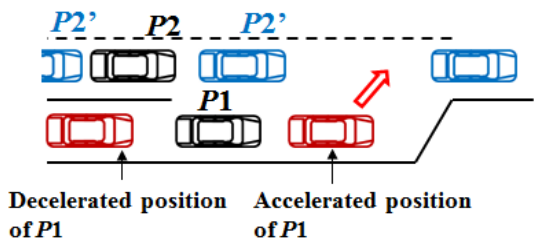

**Figure 7. The merging situation and alternate competing vehicles.**

Next, we formulate two parallel games for the respective purposes. A *merging game*, determines the merging point of the ramp vehicle based on the *current* information concerning the competing vehicles as depicted earlier in Figure 6. If the decision is to merge, then the ramp vehicle proceeds accordingly. However, if the decision is not to merge immediately, the ramp vehicle must decide whether it should accelerate or decelerate to achieve its objective of merging into the mainline. The acceleration less than 0.1 *g* does not cause passengers to feel discomfort and can be maximally 0.3 *g* for normal operation [59]. The prediced future positions of the respective vehicles (in conjunction with the payoffs associated with the various strategies) determine which vehicle in the mainline would be the co-participant in the eventual merging game as depicted in Figure 7.

In either case, the merging vehicle, *P*1, considers a 2-player

Stackelberg game with *P*2, i.e. the vehicle in the mainline[5], which may itself be simultaneously involved in another game with preceding vehicles in its adjacent lanes [21]. This interaction impacts the interplay of *P*2 with the merging vehicle, *P*1, since the decision by *P*1 does depend on the behavior of the competing vehicle, which in turn depends on its interaction with other vehicles in the mainline. We do not, however, factor in a second order effect that may be involved, i.e. the merging vehicle does not consider anything more than what the merging vehicle does and *not* why it may do so. Thus, the Stackelerg game is defined as

$$Players: P1 \text{ and } P2$$

$$Strategy\ space: S = \Gamma_1 \times \Gamma_2$$

$$\Gamma_{1,2} = \{L, S\}$$

where $\Gamma_{1,2} = \{L, S\}$ is the action set of going left, *L*, and going straight, *S*. Going left or going straight may be understood as merging in or staying in the merge lane, respectively, for the merging vehicle and as denying or yielding the right of way by the mainline vehicle[6].

The *acceleration/deceleration game* has the same solution candidates as the merging game: merging in or staying in the merge lane. Yet these do not literally mean immediately merging or staying in the merge lane. For example, the solution to merge in, of the game with an assumption of deceleration, means that if the vehicle decelerates, it may join the mainline. Notice that joining the mainline cannot be guaranteed but predicted to be possible. The same idea applies to the acceleration case. Thus in the example shown in Figure 7, we have two decisions: merging ahead of *P*2 or merging in against either vehicle marked as *P*2′, in which case *P*1 does not merge in immediately but will accelerate/decelerate first and then executes a merging maneuver. Consequently, at that point, *P*2′ takes the place of *P*2 in the *merging game*. Figure 8 shows the flowchart of the respective games.

### A. Utility Design

We define two utility functions, a positive utility and a negative utility as described below. Both the *merging* and *acceleration* games use the same utility functions. However, as stated previously, inputs for the respective utility functions are different in these case; they use *current* and *predicted* information respectively. These utility functions are designed to be adjusted by the drivers' disposition defined by the index $q \in [0,1]$, which reflects the aggressiveness of a given driver; $q = 0$ defines a completely cautious driver, $q = 1$ denotes a completely aggressive driver, while a nominally normal driver is ascribed an aggressiveness index of $q = .5$.

---

[4] Note that regardless of whether the merging vehicle accelerates or decelerates, the new competing vehicle is denoted by *P*2′ in the figure for simplicity albeit at the expense of some abuse of notation.

[5] We use *P*2 as the generic label to refer to the competing vehicle in the mainline from this point onwards.

[6] Note that going straight by the mainline vehicle is not an automatic denial of the right of way to the merging vehicle. Whether the merging vehicle perceives it as such depends on the outcome of the game-theoretic analysis, which as we see shortly refers to the inter-vehicular distances and the level of aggressiveness or risk-taking by the merging vehicle.

### 1. Utility due to headway

A basic assumption is that ordinarily drivers wish to have an appropriate headway ahead of them. This is both a matter of safety[7] and a matter of desire for free space to maneuver at will. The utility function associated with this idea is defined by

$$U_{pos} = \min(d_r, \alpha(q) \cdot d_v) \qquad (1.9)$$

where $d_r$ is the headway distance (distance between the subject vehicle and the vehicle ahead), $d_v$ denotes the visibility distance for a normal driver; $\alpha(q)$ modifies $d_v$ as a function of driver aggressiveness as depicted in Panel (a) of Figure 9.

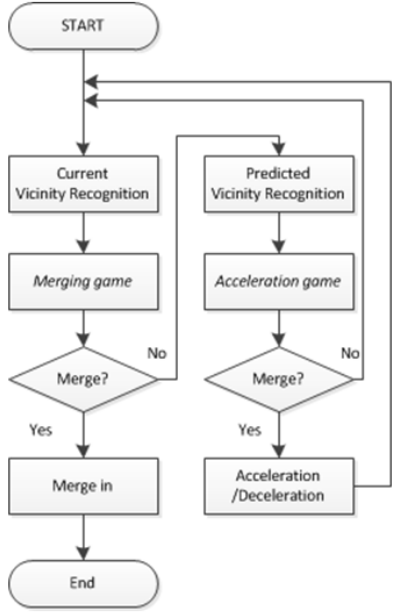

**Figure 8. Relationship between the two Stackelberg games.**

Note that this utility measure can be quantified relative to the current lane or relative to any target lane to the left or right side of the current lane as further discussed below.

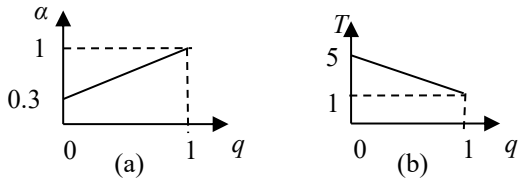

**Figure 9. Definitions of visibility range index (a) and prediction time (b) as a function of aggressiveness.**

### 2. Utility to achieve a safe lane change

There are two choices for the merging vehicle: merging-in and staying in the merge lane. Merging puts the vehicle in contention with the immediately preceding vehicle in the mainline. On the other hand, failure to merge is generally not an option as the given vehicle approaches the end of the merge lane. The utilities for the two strategies ($U_{neg,L}$ for merging or going left and $U_{neg,S}$ for staying in the same lane) are respectively defined by

---

$$U_{neg,L} = D_{suf} + v_r \cdot T(q) - d_r \qquad (1.10)$$

$$U_{neg,S} = \begin{cases} D_{suf} + v \cdot T(q) - d_e, & \text{for the merging vehicle} \\ 0, & \text{otherwise} \end{cases} \qquad (1.11)$$

where $d_r$ and $v_r$ denote the relative distance and velocity between the given vehicle and the vehicle behind in the adjacent lane, respectively. In addition, $T(q)$ designates the prediction time, $d_e$ the distance to the end of the merge lane, and $D_{suf}$ the distance essential to change lanes: e.g. an appropriate multiple of the diagonal length of the vehicle. Time headway of highway drivers [54] is used for the prediction time. The prediction time varies according to the driver's aggressiveness, $q$, as shown in Panel b of Figure 9. An aggressive driver has the minimum value among the time headway variations of highway drivers and vice versa. In general, aggressive drivers tend to change lanes or merge in even if the headway between the given vehicle and the target is comparatively small and, conversely, the prerequisite spacing to merge is larger for unaccustomed or cautious drivers.

### B. Solution of the Stackelberg Game

The solution $(\gamma^{1*}, \gamma^{2*})$ of the 2-player Stackelberg game is obtained by (1.12) and (1.14):

$$\min_{\gamma^2 \in S^2(\gamma^{1*})} U^1(\gamma^{1*}, \gamma^2) = \max_{\gamma^1 \in \Gamma^1} \min_{\gamma^2 \in S^2(\gamma^0)} U^1(\gamma^1, \gamma^2) \triangleq U^{1*} \qquad (1.13)$$

$$S^2(\gamma^1) \triangleq \{\xi \in \Gamma^2 : U^2(\gamma^1, \xi) \geq U^2(\gamma^1, \gamma^2), \forall \gamma^2 \in \Gamma^2\} \qquad (1.14)$$

where

$$U^i = U_{pos}^i - U_{neg}^i \qquad (1.15)$$

This solution is evaluated at preset periods to allow the vehicle to consider the dynamic situation that is common in driving. Note that the solution to a Stackelberg game always exists although it may not be unique [22]. The lack of uniqueness is readily remedied using the inherently safer decision should non-unique choices arise in the game theoretic analysis.

## IV. CASE STUDIES

This section presents the process of evaluation of the above game theoretic model in two merging situations. The scenarios incorporate a vehicle with a driver of selectable aggressive disposition so as to study the impact of this factor on the overall driving situation.

### A. Test Scenarios

We consider two different scenarios where there is a vehicle in the merging lane and five vehicles in the mainline as depicted in Figure 10. The merging vehicle incorporates the proposed game-theoretic decision model while the other five vehicles in the mainline move at preset speeds without changing lanes[8].

---

Each scenario is designed to provide the merging vehicle with an appropriate environment where merging is possible when the merging vehicle accelerates or decelerates respectively. We impose three different aggressiveness settings (cautious, normal, and aggressive) on the merging vehicle in each scenario. The merge entrance starts from 50 *m* in longitudinal direction and the length of the entrance is 100 *m*, with an extended 20m to complete the merge process. Lane 1, 2, and 3 and the merge lane are set to *x*=0, 3.3, 6.6, and 9.9 *m* in the lateral direction as shown in Figure 10 and Tables I and II.

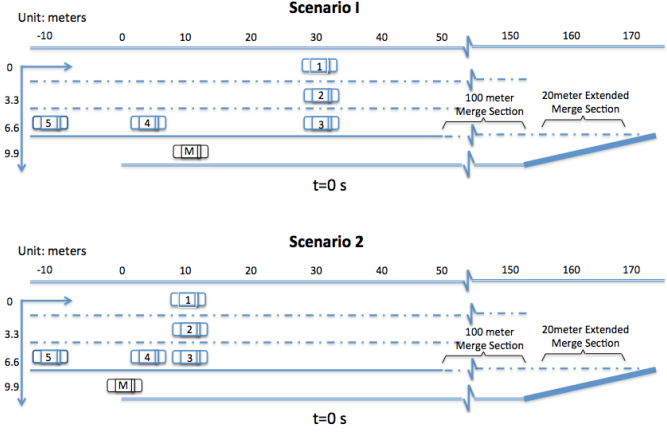

**Figure 10. Merging Scenarios.**

TABLE I.  TEST SCENARIO 1

| | Initial conditions | | |
|---|---|---|---|
| | $x_0$ (m) | $y_0$ (m) | $v_0$ (km/h) |
| Vehicle 1 in the mainline | 0 | 30 | 80 |
| Vehicle 2 in the mainline | 3.3 | 30 | 80 |
| Vehicle 3 in the mainline | 6.6 | 30 | 80 |
| Vehicle 4 in the mainline | 6.6 | 5 | 80 |
| Vehicle 5 in the mainline | 6.6 | -10 | 80 |
| Merging Vehicle | 9.9 | 10 | 70 |

TABLE II.  TEST SCENARIO 2

| | Initial conditions | | |
|---|---|---|---|
| | $x_0$ (m) | $y_0$ (m) | $v_0$ (km/h) |
| Vehicle 1 in the mainline | 0 | 10 | 80 |
| Vehicle 2 in the mainline | 3.3 | 10 | 80 |
| Vehicle 3 in the mainline | 6.6 | 10 | 80 |
| Vehicle 4 in the mainline | 6.6 | 5 | 80 |
| Vehicle 5 in the mainline | 6.6 | -10 | 80 |
| Merging Vehicle | 9.9 | 0 | 70 |

## B. Case Study Results

The results of the first scenario are depicted in Figure 11, Figure 12 and the top panel of Figure 13. The aggressive driver and the normal driver merge in while overtaking the competing vehicle (Vehicle 4) in the adjacent mainline as depicted in the bottom panel of Figure 11, whereas the cautious driver slows down to merge behind Vehicle 5 and moves further left to achieve a reasonable headway as shown in the same figure. Note that in the figure, the final position of the merging vehicle with three different aggressiveness levels (**red**, representing the aggressive vehicle, **blue**, the normal vehicle and **black**, the cautious vehicle) is superimposed on the same drawing for compactness of presentation but these cases are simulated separately.

In reference to Figure 12, it is evident that the aggressive vehicle (identified by the solid **red** line) merges quickly as its lateral position indicates whereas the normal vehicle (identified by the **blue** dashed line) merges somewhat later in the process. The cautions driver (identified by **black** dot-dashed line) merges late and subsequently moves to the middle lane as stated earlier.

The lateral vs. longitudinal positions of the vehicles studied in this scenario are depicted in the top panel of Figure 13.

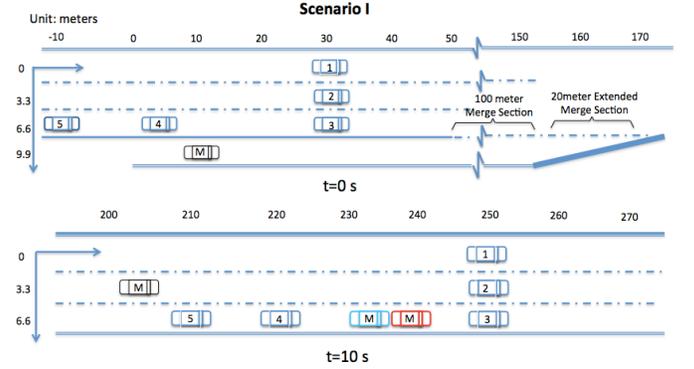

**Figure 11. Initial and Final Positions in Scenario 1.**

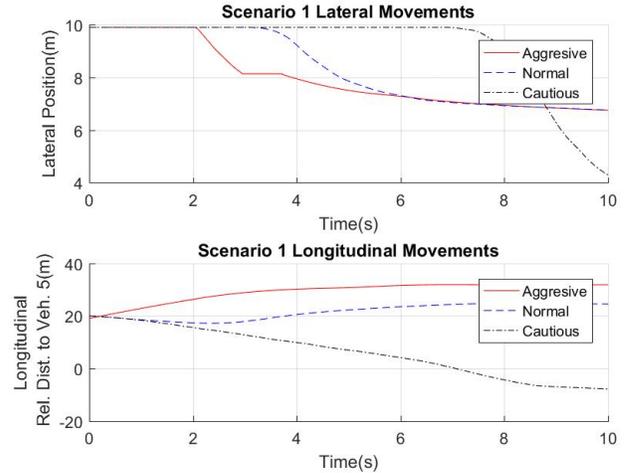

**Figure 12. The simulation results of the scenario 1.**

In above scenario, the behavior of the cautious vehicle is of some interest. It appears in both Figure 11 and Figure 12 to lag behind Vehicle 5 while merging and ends up in the middle lane subsequent to the merging process. This likely happens when the relative distance of the cautious merging vehicle with respect to Vehicle 5 is too small immediately following the merging process, causing the game theoretic model for highway driving that is programmed into the vehicle [21] to produce a discretionary lane change as depicted the aforementioned figures. The normal and aggressive vehicles, however, remain in the mainline lane where they merge since there is no advantage in shifting further to the left based on the payoffs defined in the game-theoretic model. Their longitudinal positions also reflect their behavioral dispositions in that the aggressive vehicle actually accelerates to increase its relative distance with respect to the competing mainline vehicle while the normal vehicle slows down slightly.

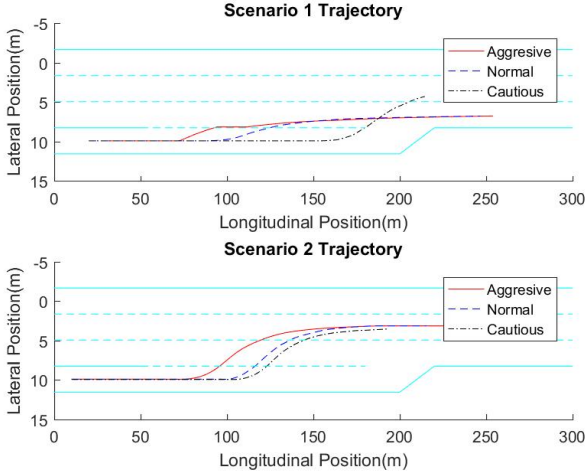

**Figure 13. Lateral vs. Longitudinal positions.**

The results of the second scenario are depicted in the bottom panel of Figure 13 and in Figure 14 and Figure 15. Note that once again, in Figure 14, the final position of the merging vehicle with three aggressiveness levels (**red**, representing the aggressive vehicle, **blue**, the normal vehicle and **black**, the cautious vehicle) is superimposed on the same drawing in bottom panel of the figure for compactness of presentation but these cases are simulated separately.

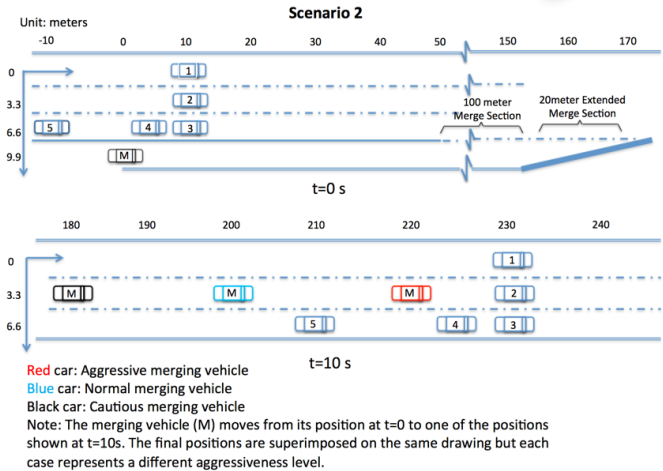

**Figure 14. Initial and Final Positions in Scenario 2.**

As the figures show, the aggressive driver will be too close behind Vehicle 4 as it merges into the rightmost lane, causing it to shift to the middle lane to improve its headway and thus positions itself behind Vehicle 2. The normal vehicle merges slightly later in the process, and would be too close behind

Vehicle 5 as it does. It too seeks a better headway by switching to the middle lane. The cautious vehicle merges later than all others and appears to be safely behind Vehicle 5 at that point. However, there is still some advantage in performing a discretionary lane change [21] to the middle lane since it would achieve an even better headway as it is evident in the respective figures.

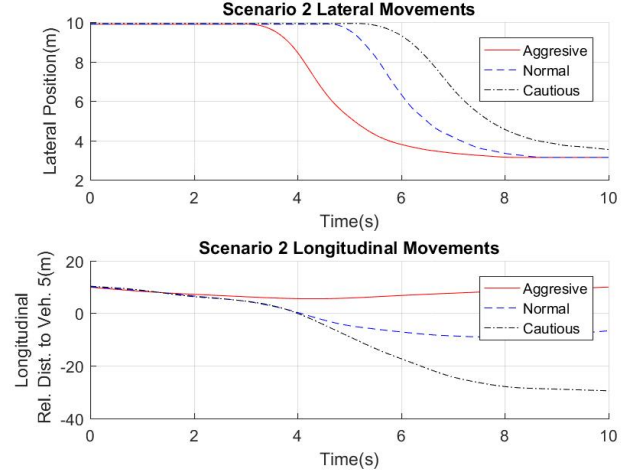

**Figure 15. The simulation results of Scenario 2**

## V. INFLUENCES ON THE MAINLINE TRAFFIC

In this section, we analyze the effects of merging on the mainline traffic. In the previous case studies, the focus was on the behavior of the merging vehicle while the vehicles in the mainline followed their respective lanes at preset speeds. In this section, we place a vehicle incorporating a game-theoretic driving model of its own [21] in the adjacent mainline. This is to investigate the resulting disturbances to the mainline traffic as a function of the aggressiveness combinations of *both* the merging-in vehicle and the vehicle in the adjacent mainline as described below.

### A. Disturbance Types

We consider two types of disturbances in the mainline traffic flow: *longitudinal* and *lateral*. The *longitudinal* disturbance, $D_{long}$, is defined as the integration of the decelerated velocity of the vehicle in the adjacent mainline, which is brought about while the merging vehicle is entering the mainline.

$$D_{long} = \int \max(v_0 - v, 0) \cdot dt \qquad (1.16)$$

Even though the mainline vehicle can speed up following the completion of the merging process to follow the merging vehicle, the amount of pullback causes upstream vehicles to slow down as well, adversely affecting the mainline flow. The *lateral* disturbance is defined as the lateral movement due to the lane-changes of the vehicle in the adjacent mainline. This is to consider the possibility that the adjacent mainline vehicle may have to execute a lane change of its own to avoid collision with the merging vehicle. Such a lane change does not necessarily result in a slowing down of the mainline vehicle but the resulting lane-changes still impacts the upstream traffic since the vehicles in the other lanes may need to slow down to accommodate the respective vehicle.

### B. Longitudinal and lateral disturbances of the mainline

Figure 16 and Figure 17 show the longitudinal and lateral disturbances in the mainline respectively, which are caused by the merging process. In particular, if the vehicle in the mainline is *highly cautious* (0% aggressiveness), the disturbances are minimized in both lateral and longitudinal directions when the merging vehicle maintains a normal aggressiveness level. This is intuitively reasonable in that the upstream traffic is not any worse off when a cautious mainline vehicle, ever so cautiously, accommodates a merging vehicle of normal aggressiveness since this behavior is *already* factored in the nominal upstream traffic flow.

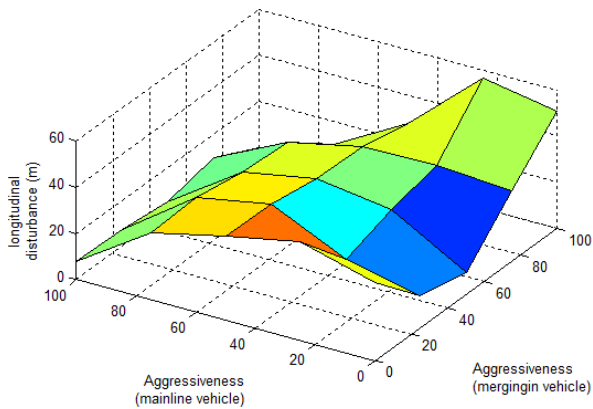

**Figure 16. Longitudinal disturbance in the mainline.**

However, there appears to be a significant impact on the mainline flow if the merging vehicle is either very cautious or highly aggressive. This is due to the fact that a cautious mainline vehicle may slow down considerably to allow an overly cautious merging vehicle to merge and similarly react to a very aggressive vehicle, which merges quickly causing the mainline vehicle to slow down measurably or to change lanes or both, particularly since an aggressive merging vehicle may merge in a manner that would lead to very small headway for the mainline vehicle regardless of the latter's disposition.

If the vehicle in the mainline maintains a *normal* aggressiveness level (50% aggressiveness), the resulting longitudinal disturbance to the mainline traffic remains at a moderate level regardless of the mode of behavior of the merging vehicle; i.e. the normal vehicle may need to slow down some to accommodate *any* merging vehicle. On the other hand, the lateral disturbance is minimized when the merging vehicle is also normal, but increases considerably with either an increase or decrease in the aggressiveness of the merging vehicle. This is understandable given the impact that a very cautious merging vehicle or a very aggressive merging vehicle may have on the mainline traffic, causing a mainline vehicle of normal aggressiveness to potentially change lanes to accommodate either of these case. In summary, a normal mainline vehicle either slows down somewhat to accommodate any merging vehicle (as one might intuitively expect) or else changes lanes to deal with highly cautious or aggressive merging vehicles. In either case, there is an overall longitudinal disturbance that propagates through the upstream traffic.

When the mainline vehicle is highly *aggressive,* it typically forces the merging vehicle to slow down and passes by the merging vehicle, or else changes lanes to accommodate a normal to aggressive vehicle. In either case, the merging vehicle typically designates a new competing vehicle relative to which it must maneuver to be able to merge into the mainline traffic. In that case, and unless the traffic density is high (which is not considered in this study), there would be sufficient spacing for the merging vehicle to safely complete its task.

The resulting longitudinal disturbance is therefore relatively modest but the lateral disturbance can be significant as depicted below. The interesting case may be the interaction between an aggressive mainline vehicle and a normal merging vehicle, which accommodates the mainline vehicle and merges without either significantly disturbing the mainline traffic in the longitudinal or lateral directions.

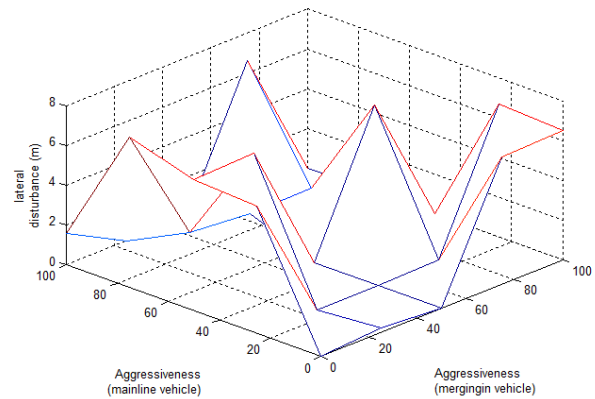

**Figure 17. Lateral disturbance in the mainline.**

### VI. CONCLUSION

In this paper, we considered a merging model based on the 2-person Stackelberg game. We designed two different games for the merging vehicle to determine the merging point and acceleration/deceleration strategy. The results indicate that merging vehicles of different levels of aggressiveness do in fact accomplish their merging objective albeit in different ways (some involving additional discretionary lane changes).

We further combined the proposed merging model with the previously developed highway driving model for discretionary lane changes to further investigate the effects of merging on the mainline traffic. When the vehicle in the mainline is *highly cautious*, both lateral and longitudinal disturbances are minimized when the merging vehicle maintains a normal aggressiveness level. However, there appears to be a significant impact on the mainline traffic if the merging vehicle is either very cautious or highly aggressive. When the vehicle in the mainline maintains a *normal* aggressive level, the resulting longitudinal disturbance to the mainline traffic remains at a moderate level, regardless of the mode of behavior of the merging vehicle. Finally, when the mainline vehicle is highly *aggressive*, the resulting and immediate longitudinal disturbance is relatively modest but the lateral disturbance can be significant. In all, the results appear to be intuitive even if in certain cases, the intuition may not fully capture the detailed behavior manifested in the scenarios presented here.

*Transportation Systems,* vol. 10, pp. 42-46, Mar 2009.

**Jehong Yoo** completed his B.Sc. and M.S. degrees in Korea and received his PhD in Mechanical Engineering from Texas A&M University in August 2014.

**Reza Langari** received the B.Sc., M.Sc., and Ph.D. degrees in mechanical engineering from the University of California, Berkeley, in 1981, 1983, and 1991, respectively. Currently, he is a professor with the Department of Mechanical Engineering, Texas A&M University, College Station.